# Analyse fonctionnelle de l'outil de gestion de planning *LibStaffer*

Résumé

*LibStaffer* est un outil de gestion de planning de service pour les bibliothèques proposé par Springshare. Dans cet article, nous exposons l'analyse que nous avons mise en œuvre pour déterminer le potentiel de *LibStaffer* en matière de simplification de la gestion de planning d'accueil, sur des implantations de configurations diverses, ainsi que d'activités transversales de service au public. Nous nous sommes fondés sur les questions établies par Philippe Lenepveu et Marc Maisonneuve, dans leur ouvrage consacré aux logiciels de gestion de planning pour les bibliothèques, et les avons enrichies par les interrogations de plusieurs collègues. Bénéficier d'un test de *LibStaffer*, de près de deux mois, nous a permis de répondre à ces questions et nous a convaincus de prendre une souscription à l'outil.

Samenvatting

*LibStaffer*, ontwikkeld door Springshare, is een planningstool om de dienstverleningen binnen bibliotheken te organiseren. Na een testperiode van twee maanden maakten we een analyse. We gingen na in welke mate *LibStaffer* bruikbaar is voor het vereenvoudigen van het beheer van de ontvangstplanning, in verschillende bibliotheekinrichtingen en voor transversale activiteiten m.b.t. de publieke dienstverlening. We baseerden ons hiervoor op de vragen die Philippe Lenepveu en Marc Maisonneuve in hun boek over planningssoftware voor bibliotheken, verrijkt met vragen van verschillende collega's. De antwoorden hierop, samen met de testresultaten, overtuigden ons om de tool in te tekenen.

--------------------------------------

**Introduction**

Assurer la gestion et le suivi des plannings des équipes de services au public ne constitue sans doute pas l'une des tâches les plus passionnantes dans le monde des bibliothèques. Outre le fait que cette activité peut se montrer particulièrement chronophage, elle implique couramment une complexe distribution pour le personnel qui exerce sur plus d'un site, avec des compétences diverses, des statuts différents, des sensibilités propres, des attentes spécifiques, sans oublier les éventuels étudiants-jobistes à accorder à la vie des équipes. Pour parvenir à gérer ces contraintes, un logiciel de gestion de planning de service peut se révéler des plus utiles.

**Contexte**

En 2018, la Bibliothèque de l'Université de Liège (ULiège Library) a entamé une profonde refonte de son organisation. Plusieurs entités et services semi-autonomes ont ainsi été fusionnés au sein d'une nouvelle structure renforcée, ULiège Library, constituée d'environ 100 personnes. La nouvelle disposition d'ULiège Library rendait davantage opportun le besoin en outils communs, notamment pour une gestion au quotidien plus efficace et transparente[1]. Un point fréquemment abordé lors des débats participatifs de début 2021 (dans le cadre de la réforme) visait la problématique de la combinaison de l'accueil des usagers, à assurer sur la quinzaine d'implantations de bibliothèques, avec les activités transversales, c'est-à-dire non liées à un site, à des usagers spécifiques ou à une discipline en particulier, et réalisées au profit de tout utilisateur d'ULiège Library (prêt interbibliothèques, support en ligne aux usagers, service des

---

[1] Ishak, Yuyun Wirawati. Technology that enables, services that empower. *IFLA WLIC 2018: Reference and Information Services joint with Information Technology Satellite, Kuala Lumpur, Malaysia, August 23* [en ligne], 2018 (consulté le 2 mars 2022). <https://ink.library.smu.edu.sg/library_research/128>.

acquisitions…). À côté de la mise en place de dispositifs de communication interne plus efficaces à l'échelle d'une structure unifiée, se doter d'un outil professionnel dédié à la gestion de ces plannings semblait une étape essentielle.

En 2019, Philippe Lenepveu et Marc Maisonneuve ont édité une étude comparative[2] des logiciels de gestion de planning de service pour les bibliothèques. Trois solutions y ont été présentées en détail : *Bcal*, *Credo-Planning* et *Planning Biblio*. La solution logicielle sous la forme de service (*Software as a Service,* SaaS) *LibStaffer*, de la société Springshare, n'intégrait pas l'analyse. Les auteurs de l'ouvrage se sont en effet volontairement limités aux outils qui disposaient d'un bureau ou d'un point de contact en France. ULiège Library disposant déjà d'une souscription à un produit de la société Springshare (*LibCal*), il nous a semblé judicieux d'analyser le fonctionnement de *LibStaffer* en exploitant la grille d'analyse des auteurs de l'étude.

**Revue de la littérature**
Les publications spécifiquement relatives à *LibStaffer* ne foisonnent pas. Ellie Dworak établit un premier état des lieux des fonctionnalités de l'outil, en 2014[3]. Pour des informations à jour, le lecteur s'orientera en priorité vers le portail d'aide *Get help with LibStaffer* de Springshare, fréquemment mis à jour et dont une partie est accessible en mode invité. Parmi les autres outils de gestion de planning utilisés en bibliothèque, nous épinglons aussi *Teamup*[4] et *Google*[5]. Le lecteur intéressé par la gestion de planning spécifiquement en lien avec des étudiants-jobistes se tournera en priorité vers les articles de Kindra Becker-Redd, Kirsten Lee et Caroline Skelton, paru en 2018[6], de Janiece Jankowski paru en 2013[7] et de Bixia H. Tang paru en 2020[8].

**Méthodologie**
Nous sommes partis de la soixantaine de questions de nature fonctionnelle relatives aux outils de gestion de planning qui figurent dans l'ouvrage de Lenepveu et Maisonneuve. La répartition en grandes rubriques de questions a également été conservée. Les questions ont été soumises fin juin 2021 aux différents coordinateurs de terrain des cinq plus grands sites de ULiège Library. Il leur a été demandé de prendre connaissance des questions sur lesquelles le travail d'analyse allait se baser, de signaler pour la mi-août si certains points méritaient une attention particulière (par exemple, "à la question X, être particulièrement vigilant à Y") et si des questions et éléments d'analyse méritaient d'être ajoutés. Ils ont en outre été invités à partager la liste avec des collègues impliqués au quotidien dans la gestion du planning des équipes.

---

[2] Lenepveu, Philippe ; Maisonneuve, Marc. *Les logiciels de gestion de planning de service pour les bibliothèques*. Éditions Klog, 2019. ISBN 979-10-92272-17-8.
[3] Dworak, Ellie. LibStaffer from Springshare. *The Charleston Advisor* [en ligne], January 2014 (consulté le 2 mars 2022), vol. 15, no 3, p. 32-35. <https://doi.org/10.5260/chara.15.3.32>
[4] Hughes, Sarah E. Scheduling Using a Web-Based Calendar: How Teamup Enhances Communication. *Public Services Quarterly* [en ligne], 2018 (consulté le 2 mars 2022), vol. 14, no 4, p. 362-372 <https://doi.org/10.1080/15228959.2018.1518184>.
[5] Tang, Bixia H. Inviting student employees to work – a scheduling system built with free google applications, *Public Services Quarterly* [en ligne], 2020 (consulté le 2 mars 2022), vol. 16, no 4, p. 254-264. <https://doi.org/10.1080/15228959.2020.1806178>.
[6] Becker-Redd, Kindra ; Lee, Kirsten ; Skelton, Caroline. Training Student Workers for Cross-Departmental Success in an Academic Library: A New Model, *Journal of Library Administration* [en ligne], 2018 (consulté le 2 mars 2022), vol. 58, no 2, p. 153-165. <https://doi.org/10.1080/01930826.2017.1412711>.
[7] Jankowski, Janiece. Successful Implementation of Six Sigma to Schedule Student Staffing for Circulation Service Desks, *Journal of Access Services* [en ligne], 2013 (consulté le 2 mars 2022), vol. 10, no 4, p. 197-216. <https://doi.org/10.1080/15367967.2013.830930>
[8] Voir note 5.

Plusieurs interrogations et remarques ont été adressées et celles-ci ont permis de compléter la grille d'analyse initiale avec une vingtaine de nouvelles questions. Afin de les distinguer de celles de l'ouvrage, ces questions issues de l'équipe de ULiège Library sont précédées d'un astérisque dans les différentes rubriques concernées.

Début septembre 2021, une période d'essai de *LibStaffer* a été demandée à la société Springshare. Une extension a été sollicitée pour le mois d'octobre. Le test a ainsi pu être poursuivi jusqu'au 27 octobre et nous avons été en mesure de répondre à l'ensemble des questions listées. En outre, comme dans l'ouvrage de Lenepveu et Maisonneuve, de nombreuses notes et explications complémentaires ont été apportées car, fréquemment, une simple réponse oui/non s'avérait réductrice et insignifiante pour appréhender valablement les potentialités et limites de l'outil.

Enfin, afin de tester au mieux *LibStaffer* dans un contexte concret, des données réelles et des éléments de configuration relatifs à des implantations de ULiège Library (Santé-CHU, Polytech, L. Graulich et la salle de lecture principale) ont été exploités ("qui travaille quand et où ?"). Ceux-ci ont permis de réaliser une configuration de base, de construire des modèles, de vérifier les règles de contrôle, de tester les droits des utilisateurs, etc. et ainsi de mettre à l'épreuve le système jusque dans le moindre détail, opérations que Lenepveu et Maisonneuve n'ont, semble-t-il, pas eu l'opportunité de réaliser dans leur étude comparative.

**Analyse de la solution**
   1) Les horaires de la bibliothèque

| Question | Réponse | Note |
|---|---|---|
| Le logiciel permet-il de définir les horaires hebdomadaires d'ouverture de la bibliothèque ? | Oui | a |
| Le logiciel permet-il de définir des horaires différents pour différentes périodes de l'année ? | Oui | b |
| Le logiciel permet-il d'établir le calendrier annuel des heures d'ouverture de la bibliothèque ? | Oui | |
| Le logiciel permet-il de calculer le nombre d'heures d'ouverture dans l'année et par période de temps ? | Oui | c |

   a. Il convient toutefois de nuancer cette assertion. Les horaires sont synchronisables avec les horaires de la solution *LibCal* (Springshare) (outil qui permet de gérer des calendriers d'événements et de manifestations, la réservation de salles de travail, de places en bibliothèque et d'équipements particuliers), dans la mesure où les horaires de *LibCal* peuvent eux-mêmes être synchronisés avec le système de gestion de bibliothèque Alma. À la création d'un quart de travail (*shift*)[9], les heures d'ouverture de la bibliothèque sont indiquées, mais cela n'empêche pas de créer un quart en dehors des heures d'ouverture.

---

[9] *LibStaffer* a été testé dans ses interfaces anglaise et française. Dans un souci de clarté, certaines fonctionnalités et caractéristiques de l'outil sont mentionnées en français avec le pendant anglais de l'interface d'origine.

Une tâche en dehors des heures d'ouverture et d'accès de la bibliothèque peut ainsi être prévue.
b. Oui, grâce à la synchronisation avec *LibCal*. Dans les faits, disposer ou non d'horaires de la bibliothèque n'a aucun impact si ce n'est la possibilité de superposer l'horaire aux quarts de travail créés. In fine, *LibStaffer* n'est pas un outil destiné à gérer les horaires de la bibliothèque.
c. En passant par la fonctionnalité "Rapports" et en filtrant sur les colonnes du tableau Excel qui peut être généré à partir d'un export en csv.

2) Les horaires des services

| Question | Réponse | Note |
|---|---|---|
| Pour chaque semaine type (correspondant à un horaire d'ouverture hebdomadaire de la bibliothèque), le logiciel permet-il de définir les horaires de chaque service ? | Oui | |
| Le logiciel permet-il de calculer le nombre d'heures d'ouverture de chaque service dans l'année ou par période de temps ? | Oui | a |

a. En passant par "Rapports" et en filtrant sur les colonnes du tableau Excel qui peut être généré à partir d'un export en csv.

3) La prévision des charges de service et des postes de travail

| Question | Réponse | Note |
|---|---|---|
| Pour chaque service, le logiciel permet-il de préciser, par tranche horaire, les ressources nécessaires (le nombre de postes ou le nombre d'agents) pour assurer le service ? | Oui | a |
| Le logiciel permet-il d'obtenir un état des ressources nécessaires (en nombre d'heures par semaine) pour assurer chacun des services ? | Non | |

a. Par défaut, le système estime qu'il faut une personne par quart de travail. Le gestionnaire du planning (*schedule*) peut augmenter cette valeur ou la porter à zéro. Un quart de travail qui serait en sous-effectif est mis en évidence avec des hachures de couleurs.

4) La gestion des ressources

La gestion des agents

| Question | Réponse | Note |
|---|---|---|
| Le logiciel permet-il de renseigner les informations suivantes : | | |
| Le nom de l'agent ? | Oui | |
| Le statut de l'agent (par exemple « interne » pour un agent de la bibliothèque ou « externe » pour un intervenant extérieur) ? | Oui | a |

| | Réponse | Note |
|---|---|---|
| Le nom de l'entreprise ou de l'association dans le cas d'un intervenant extérieur ? | Oui | b |
| La catégorie administrative de l'agent (A, B, C) ? | Oui | a |
| Le temps de travail hebdomadaire ? | Oui | |
| Le quota d'heures de service public à effectuer ? | Oui | c |
| *Le cas échéant, le nombre de catégories est-il limité ? | Non | |
| *Le cas échéant, les catégories peuvent-elles être différenciées, notamment via des couleurs ? | Oui | d |

a. Cette information peut être renseignée via la fonctionnalité "Positions". Chaque compte agent dans *LibStaffer* peut être associé à une ou plusieurs positions (ex : documentaliste, étudiant-jobiste, magasinier, agent d'accueil, responsable, stagiaire…), définies par l'administrateur du système. L'assignation d'un quart de travail à un agent peut être conditionnée dans un planning à la présence d'une position particulière. Ainsi, certains quarts de travail peuvent être réservés à du personnel spécifique.
b. Il existe plusieurs possibilités. Ce type d'information pourrait être renseigné :
    - soit dans la zone même du nom ou du prénom de la personne ;
    - soit dans la zone Département, laquelle est une zone contrôlée dont les valeurs sont préalablement créées par un administrateur *LibStaffer* ;
    - soit via *LibApps > Admin > Manage Account* via l'onglet "Profile Box". Une exploitation rapide de l'information (par exemple sous forme de listing) ou à visée fonctionnelle dans *LibStaffer* n'est cependant pas prévue ;
    - soit via les "Positions" qui mentionneraient spécifiquement un nom d'entreprise ou d'association tierce.
c. Il existe plusieurs quotas qui sont définissables au niveau de chaque agent :
    - nombre maximal d'heures consécutives prestées
    - nombre maximal de jours consécutifs prestés
    - nombre maximal d'heures par jour
    - nombre minimal d'heures par semaine
    - nombre maximal d'heures par semaine
    - nombre maximal d'heures par mois
d. Chaque agent renseigné peut se voir attribuer une couleur précise via un code hexadécimal. L'administrateur peut aussi appliquer la même couleur à tous les comptes avec une position particulière (ex. les étudiants jobistes). Cela peut ainsi aider à différencier certaines catégories de personnel.

<u>La gestion des habilitations des agents et des objectifs de service</u>

| Question | Réponse | Note |
|---|---|---|
| Le logiciel permet-il de préciser les services que chaque agent peut assurer ? | Oui | a |
| Pour chaque agent, le logiciel permet-il de préciser un quota d'heures pour chacun des services qu'il peut assurer ? | Non | b |
| Le logiciel permet-il de définir d'autres priorités d'affectation ? | Oui | c |

  a. Plusieurs départements peuvent être attribués à un agent. Il est par exemple possible d'assigner deux départements comme "Accueil" et "Prêt interbibliothèques". De plus, les différents plannings peuvent s'envisager comme élaborés autour des services à proprement parler (Accueil Bibliothèque ABC, Accueil Bibliothèque DEF, Prêt interbibliothèques, Service de référence virtuelle, etc.) et un même agent peut être associé à plusieurs de ces plannings.
  b. Il est possible de définir un quota d'heures total (voir ci-dessus), et non par service (planning). Via le planificateur automatique (*Auto Scheduler*), il est en outre possible de limiter par agent le nombre de quarts de travail sur une journée (voir plus bas).
  c. Des quarts de travail favoris peuvent être renseignés. Ces favoris sont des indicateurs qui peuvent être utilisés par le gestionnaire du planning ou exploités par l'outil du planificateur automatique (voir ci-dessous).

La gestion de la disponibilité des agents

| Question | Réponse | Note |
|---|---|---|
| Le logiciel permet-il de gérer le planning hebdomadaire de chaque agent afin d'obtenir automatiquement un état de présence (ou un état des disponibilités) pour chaque jour de l'année ? | Oui | |
| Le logiciel permet-il de gérer les périodes d'absence des agents (les périodes de congés, les périodes de formation, etc.) ? | Oui | a, b |
| Le logiciel permet-il d'enregistrer les absences ou les indisponibilités exceptionnelles des agents ? | Oui | a |
| Un agent peut-il lui-même signaler ou enregistrer une absence ou une indisponibilité ? | Oui | c |
| En cas d'absence ou d'indisponibilité signalées par un agent, le logiciel permet-il d'envoyer une alerte au gestionnaire de planning ? | Oui | |
| *En cas d'absence ou d'indisponibilité d'un agent, le logiciel permet-il de repérer quels sont les autres agents disponibles pour le remplacement ? | Oui | |
| *Le logiciel permet-il aux agents de s'inscrire volontairement dans des plages de service ? | Oui | d |
| *Le logiciel permet-il aux agents de procéder à des échanges de service ? | Oui | e |

  a. Si un quart de travail a déjà été attribué à un agent et que celui-ci est indiqué absent (*Time Off*) par la suite, le quart lui reste attribué. Par contre, l'agent ne sera pas proposé pour l'attribution d'un quart après qu'il a été marqué comme indisponible. Il est regrettable que cela ne soit pas plus visible dans le planning. Même si un nombre minimum d'agents est indiqué pour un shift, si un des agents est en Time off et que le nombre minimum n'est plus atteint, le shift ne devient pas hachuré et donc aucune alerte bien visible n'est affichée.
b. Lorsqu'un agent est par exemple en formation, si cela est indiqué via la fonctionnalité *Time Off*, il ne sera pas renseigné comme éventuellement disponible pour être assigné à un quart de travail. Si nécessaire, le gestionnaire du calendrier peut toutefois forcer une assignation en cas d'indisponibilité. Par exemple, si l'agent est habituellement en indisponibilité en raison d'un cours d'anglais hebdomadaire suivi durant ses heures de travail habituelles, le gestionnaire du calendrier peut forcer l'assignation à un quart de travail pour effectuer l'accueil en bibliothèque et supprimer par la suite l'indisponibilité existante. La liste des moments d'indisponibilité de tous les agents d'un même planning est visible de ces mêmes agents, quelle que soit leur position. Cette transparence de l'information peut s'avérer très pratique pour les plannings ou bibliothèques dont l'équipe est relativement importante.
c. Permettre ou non aux agents de renseigner leurs indisponibilités via la fonctionnalité des *Time Off* est une option qui peut être activée ou désactivée par un administrateur. L'option choisie est valable pour l'ensemble de la solution *LibStaffer* et s'applique à tous les utilisateurs réguliers (voir ci-dessous), quelle que soit leur position. Une demande d'indisponibilité introduite par un agent peut également être soumise à validation à son/sa responsable hiérarchique et/ou aux administrateurs. Cette fonctionnalité est également à activer/désactiver au niveau général dans la configuration du système *LibStaffer*. Selon la configuration établie, après une assignation à un quart, un agent peut en outre s'en défaire via les fonctionnalités "Abandonner" (*Give up*), à charge pour lui qu'un collègue automatiquement alerté accepte de le remplacer, ou "Laissez tomber" [sic] (*Drop*), à charge pour le gestionnaire de lui trouver un remplaçant si l'agent ne désigne pas lui-même la personne qui reprendra son quart de travail.
d. Via la fonctionnalité "Réclamation de quart" (*Shift Claiming*), un utilisateur peut prendre des quarts de travail de son planning pour lesquels le nombre de ressources humaines nécessaires ne serait pas encore atteint. Si la fonctionnalité est activée, un utilisateur peut aussi échanger un quart avec un.e collègue disponible au sein du même planning.
e. Selon la configuration choisie, avec ou sans validation par le gestionnaire du planning (fonctionnalité "Échanger" [*Swap*]). À noter également que grâce à la fonctionnalité "Diviser" (*Split*), un quart de travail peut être scindé en entités plus réduites, offrant ainsi plus de souplesse pour des possibilités de *Give up*, *Swap* et *Drop*[10].

5) L'élaboration du planning de service

| Question | Réponse | Note |
|---|---|---|
| Le logiciel permet-il d'élaborer différentes grilles horaires de service pour chaque jour de la semaine et chaque semaine de l'année ? | Oui | |
| Le logiciel permet-il de définir librement les créneaux horaires pour chaque poste de travail ? | Oui | |
| Si oui, avec quelle précision (par tranche d'une heure, d'une demi-heure, d'un quart d'heure, de cinq minutes) ? | Oui | a |
| Lors de l'affectation d'un agent à un poste de travail, le logiciel fournit-il une aide (par exemple sous la forme d'une liste déroulante) | Oui | b |

---

[10] L'existence des fonctionnalités *Split* et *Drop* est ultérieure à l'analyse de Dworak (2014).

| | | |
|---|---|---|
| afin qu'on ne puisse pas affecter un agent à un poste s'il n'est pas habilité à occuper ce poste ou s'il n'est pas disponible ? | | |
| Le logiciel effectue-t-il des contrôles d'intégrité afin d'empêcher qu'un agent soit affecté simultanément à deux postes dans le même créneau horaire de la même journée ? | Oui | c |
| *Au sein d'un même planning de service, le logiciel permet-il d'assigner simultanément et de façon récurrente un même agent à deux postes dans le même créneau horaire de la même journée (par exemple pour assigner en même temps une tâche principale [accueil…] et une tâche secondaire [monitoring d'une boîte email…]) ? | Non | d |
| *Au sein de plannings de service différents, le logiciel permet-il d'assigner simultanément et de façon récurrente un même agent à deux postes dans le même créneau horaire de la même journée (par exemple accueil en implantation peu fréquentée dans un planning et gestion du service de référence virtuel ou service de chat dans l'autre planning) ? | Non | |
| Le logiciel affiche-t-il une alerte en cas de contradiction avec des règles de gestion prédéfinies (pause déjeuner, durée de service trop longue…) ? | Oui | e |
| Lors de l'affectation d'un agent à un poste de travail, le logiciel fournit-il une aide au choix (par exemple sous la forme d'une liste triée) afin d'affecter en priorité les agents selon certains critères (comme le nombre d'heures déjà effectuées ou restant à effectuer) ? | Oui | f |
| Le logiciel dispose-t-il d'une fonction pour mettre en évidence (ou en « surbrillance ») les heures de service d'un agent particulier ? | Oui | g |
| Afin de répartir au mieux les charges de service, le logiciel permet-il, en cours de saisie, de contrôler à tout moment et pour chaque agent (par exemple sous la forme d'un tableau de bord), le quota des heures effectuées et restant à effectuer (total des heures de service public, détail des heures par service…) ? | Oui | h |
| *Le logiciel propose-t-il un moyen de tenir compte des heures supplémentaires dans le calcul global des heures prestées ? | Oui | i |
| *Le logiciel propose-t-il un moyen de tenir compte des heures particulières (prestations en soirée, le week-end…) dans le calcul global des heures prestées ? | Non | |
| *Le logiciel propose-t-il un moyen de tenir compte des heures prestées durant les dispenses de service octroyées par l'institution ? | Non | |
| *Le logiciel permet-il de gérer les éventuelles prestations en travail à domicile ? | Oui | j |

a.  Les quarts de travail peuvent être définis si nécessaire à la minute près.
b.  Pour chaque quart de travail, la (les) position(s) requise(s) pour pouvoir occuper le quart peut/peuvent être indiquée(s). Seuls les utilisateurs habilités et appartenant au planning en question peuvent être sélectionnés.
c.  Le gestionnaire conserve toutefois la possibilité d'outrepasser le conflit au cas par cas (c'est-à-dire de manière non récurrente au sein du quart planifié).
d.  Par contre, il est tout à fait possible de créer dans le planning des quarts de travail couvrant plusieurs tâches[11]. Il existe aussi une solution de contournement qui consiste à créer des quarts pour certaines tâches en dehors des heures de service normales. Par exemple, le relevé de la boîte email de la bibliothèque peut être assigné dans *LibStaffer* à un agent pour la période allant de 7 à 8 heures du matin (moment où la bibliothèque est encore fermée), ce même agent assurant l'accueil de 8 à 12 heures. Dans les faits, ces deux tâches seront accomplies dans le même temps, durant toute la présence de l'agent à l'accueil. Il est ainsi possible d'assigner à un même agent, de façon récurrente, des quarts de travail relatifs à des missions différentes, mais qui concrètement seront effectuées simultanément. Cette approche n'est pas des plus élégantes, mais satisfait aux cas de réelle nécessité.
e.  Il faut paramétrer précisément les comptes de chaque utilisateur (onglets "Gérer le compte" [*Manage Account*] et "Gérer les heures disponibles" [*Manage Available Hours*]) pour, par exemple, empêcher toute planification durant un temps de midi. Cette configuration peut être réalisée par l'agent lui-même, voire par un administrateur du système. Il est à noter toutefois que ce planning de disponibilité générale n'existe qu'au niveau du compte de l'agent et ne tient pas compte par exemple de la prestation de mi-temps dans deux implantations de bibliothèque différentes. Ainsi, il n'est pas possible de préciser via "Gérer les heures disponibles" que l'agent X travaille dans l'implantation A le lundi et le mardi et dans l'implantation B du mercredi au vendredi. Cette distinction n'est gérée actuellement qu'au niveau du planning[12].
    En ce qui concerne une éventuelle durée de service trop longue, celle-ci n'est pas configurable en tant que telle, mais les durées des quarts sont définies pour chaque planning et il est possible de donner une valeur limite du nombre d'heures consécutives prestées au niveau du compte de chaque utilisateur (voir ci-dessus).
f.  Cette possibilité est offerte par l'outil du planificateur automatique qui assigne automatiquement les utilisateurs d'un planning aux quarts en sous-effectif. Via le planificateur, c'est tout le planning qui sera traité, éventuellement restreint sur la base de plages de dates, et non certains quarts seulement qui seraient sélectionnés. Si nécessaire, on peut par exemple limiter de 1 à X quarts assignés par jour et déterminer qu'un utilisateur aura Y minutes ou Z heures entre deux quarts assignés.
    Au niveau de chaque quart, un ou plusieurs utilisateurs favoris peuvent être déterminés pour remplir le quart. Cette donnée ne peut être utilisée dans le planificateur automatique que :
    - si l'on sélectionne "Planification automatique - Personnel favori uniquement" (*Auto Schedule Favorite Staff only*), seuls les utilisateurs marqués comme favoris pourront remplir le quart. S'ils sont tous indisponibles, le quart restera vide.

---

[11] Dworak, Ellie. LibStaffer from Springshare. *The Charleston Advisor* [en ligne], January 2014 (consulté le 2 mars 2022), vol. 15, nº 3, p. 32-35. <https://doi.org/10.5260/chara.15.3.32>.

[12] Cette fonctionnalité a toutefois récemment été demandée par quelques clients : "Staff working multiple locations" <https://lounge.springshare.com/discussion/663/staff-working-multiple-locations> [accès après authentification] (consulté le 2 mars 2022).

- - si l'option "Planification automatique - Personnel favori uniquement" n'est pas sélectionnée, certains quarts seront attribués à des utilisateurs non marqués comme favoris. Mais si tous les utilisateurs indiqués comme favoris sont indisponibles, le quart sera rempli avec un autre utilisateur disponible.
  g. Cette mise en évidence est possible via la vue "Chronologie du personnel" (*Staff Timeline*) et peut être renforcée par un filtre appliqué sur un utilisateur particulier ainsi que l'éventuelle couleur assignée à un utilisateur ou groupe d'utilisateurs. Pour un nombre important d'utilisateurs dans un même planning, les nuances de couleur pourraient être plus difficiles à distinguer. Une seconde possibilité est de passer par la vue "Vue multi-planification" (*Multi-Schedule View*) et d'appliquer un filtre sur l'utilisateur après avoir sélectionné les plannings dans lesquels il est affecté.
  h. Pour les administrateurs, ces informations sont accessibles via le menu "Personnel" (*Staff*) et en sélectionnant le nom de l'agent. Au sein d'un planning, chaque agent peut visualiser les quarts de ses collègues du planning ainsi que les quarts non assignés. Par contre, il n'existe pas d'accès direct à un état de la situation des affectations et des quotas (sauf à exécuter des rapports et analyses via le menu "Rapports").
  i. Le logiciel permet d'établir des calculs de taux horaire pour des heures normales et des heures supplémentaires. Cette configuration se fait au niveau de chaque utilisateur.
  j. Deux façons de faire sont ici possibles :
    - soit au moyen d'un planning dédié : une implantation ou un service peut par exemple disposer de plusieurs plannings, l'un pour le travail en présentiel, l'autre pour le travail à domicile[13]. Cette procédure peut par exemple très bien convenir à des implantations ou services avec un personnel nombreux.
    - soit dans un planning existant : pour les implantations où le travail à domicile n'est pas très fréquent ou avec peu de personnel, des quarts "Travail à domicile" peuvent être ajoutés dans le planning même, ce qui permet d'avoir une vision globale dans un seul planning.

6) La gestion des modèles

| Question | Réponse | Note |
|---|---|---|
| Le logiciel permet-il de créer des modèles de grilles de service ou permet-il de dupliquer une grille de saisie existante ? | Oui | a |
| Le logiciel permet-il d'élaborer différents modèles de grilles de service afin d'organiser au mieux la rotation du personnel (par exemple une grille pour les semaines « paires » et une autre pour les semaines « impaires », des grilles différentes pour les semaines « A », « B », « C ») ? | Oui | b |
| Lors de la création d'une nouvelle grille de service sur base d'un modèle, les informations sont-elles bien remplies avec les valeurs par défaut ? | Oui | |
| Lors de la création d'une nouvelle grille de service sur base d'un modèle, le logiciel signale-t-il les cases à corriger (compte tenu des indisponibilités des agents ce jour-là) ? | Oui | |

---

[13] Voir la vidéo *LibStaffer for Hybrid and Virtual Environments* <https://training.springshare.com/libstaffer/hybrid-and-virtual> [accès après authentification] (consulté le 2 mars 2022)

a. La fonctionnalité de copie des quarts de travail peut se faire en tenant compte du personnel assigné dans les quarts originaux (avec ou sans vérification de sa disponibilité) ou sans copie du personnel (création de quarts non assignés).
b. Cette fonctionnalité vaut pour la création des quarts de travail, mais également pour l'assignation du personnel dans les quarts.

7) La publication du planning

| Question | Réponse | Note |
|---|---|---|
| Le logiciel permet-il la consultation en ligne du planning de service de la bibliothèque ? | Oui | a |
| Chaque agent dispose-t-il d'une entrée particulière lui permettant de consulter son planning personnel ? | Oui | b |

a. Un planning peut être rendu public ou laissé accessible aux seules personnes disposant d'un compte *LibStaffer* actif. Si le planning est public, il peut être diffusé via une URL ou un code d'incorporation généré à partir d'un outil widget. *LibStaffer* peut également n'être rendu accessible qu'à certaines plages IP.
b. Chaque utilisateur dispose, en page d'accueil, d'un tableau de bord listant les quarts de travail assignés, par défaut au cours des 7 prochains jours, mais la période affichée peut être personnalisée.

8) Les statistiques

| Question | Réponse | Note |
|---|---|---|
| Le logiciel dispose-t-il d'un modèle de production de statistiques intégré ? | Oui | |
| Ce module permet-il de générer des graphiques ? | Non | |
| Le logiciel propose-t-il différents états statistiques prédéfinis pour répondre aux besoins d'analyse les plus courants ? Si oui, combien ? | Oui | |
| L'utilisateur peut-il lui-même créer ses propres requêtes et états statistiques ? | Oui | |
| Le logiciel permet-il la visualisation des statistiques dans un format imprimable et l'exportation dans un format de document « portable » (PDF…) ? | Oui | |
| Le logiciel permet-il l'exportation des donnée dans un format approprié en vue d'une analyse au moyen d'un logiciel tableur ? | Oui | a |
| *L'accès aux données statistiques est-il possible à d'autres personnes que les administrateurs ? | Oui | b |

a. Export possible en csv.

b. Chaque compte utilisateur doit être défini comme "régulier" (*Regular*) ou "administrateur" (*Admin*). Un utilisateur régulier peut disposer de droits élargis dans un planning donné et ainsi accéder aux statistiques de ce planning.

9) La gestion des plannings dans le cadre d'un réseau de bibliothèques

| Question | Réponse | Note |
|---|---|---|
| Le logiciel permet-il la gestion d'un réseau de bibliothèques (sans limite du nombre de ces bibliothèques) ? | Oui | |
| Chaque bibliothèque peut-elle gérer ses propres plannings de manière indépendante (avec des horaires différents, des contraintes de service différentes, etc.) ? | Oui | |
| Le référentiel des agents est-il commun pour l'ensemble des bibliothèques du réseau ? | Oui | |
| Un même agent peut-il être affecté au service de plusieurs bibliothèques ? | Oui | a |
| Toutes les informations sont-elles bien gérées dans la même base de données (et accessibles via la même application) ? | Oui | |
| Le logiciel permet-il de produire un état de service global (pour l'ensemble des bibliothèques) et des états de service distincts pour chaque bibliothèque (statistiques, etc.) ? | Oui | |

a. Dans la structure de *LibStaffer*, on peut définir 3 niveaux hiérarchiques[14] :
   - des localisations (*Locations*)
   - des départements (*Departments*)
   - des positions (*Positions*)

Un agent peut (mais ne doit pas) être assigné à une ou plusieurs localisations, un ou plusieurs départements et/ou une ou plusieurs positions. Les positions sont utiles au contrôle des affectations. Par exemple, si on souhaite réserver certains quarts à des agents d'accueil ou des étudiants jobistes, les département et localisations n'ont pas de rôle fonctionnel, mais peuvent être exploités dans le cadre de statistiques.

10) La gestion des utilisateurs et des droits

| Question | Réponse | Note |
|---|---|---|
| Le logiciel permet-il la connexion à un annuaire externe via le protocole LDAP (*Lightweight Directory Access Protocol*) ? | Oui | |
| Le logiciel permet-il l'authentification unique (*Single Sign-On*) via un serveur CAS (*Central Authentication Service*) ? | Oui | a |
| Le logiciel permet-il la gestion des rôles suivants : | | |

---

[14] Voir *Accounts & Org Hierarchy: Set up and assign locations, departments, and positions* <https://ask.springshare.com/libstaffer/faq/1978> (consulté le 2 mars 2022)

| | | |
|---|---|---|
| ● Administrateur de l'application ? | Oui | |
| ● Administrateur du planning ? | Oui | b |
| ● Agent de service ? | Oui | |
| Le logiciel permet-il de définir précisément les droits par groupes d'utilisateurs et pour chaque utilisateur (gestion des plannings, gestion des absences, gestion des remplacements…) ? | Oui | |
| Dans le cas d'un réseau de bibliothèques, le logiciel permet-il de définir précisément les droits d'administration et d'utilisation pour chaque bibliothèque (administrateurs, agents) ? | Oui | |
| Le logiciel comprend-il une gestion de « workflow » pour la gestion des différentes étapes de la création du planning à sa diffusion ? | Oui | c |
| *Le logiciel permet-il de vider la base de de données des anciennes activités et de purger les comptes devenus obsolètes ? | Oui | d |

a. L'identification est gérée via le module *LibAuth* automatiquement inclus dans une souscription à un produit Springshare. Outre une authentification possible directement sur base de l'adresse email et d'un mot de passe stockés dans *LibAuth*, les interfaçages avec les outils et protocoles suivants sont possibles : SAML (y compris Shibboleth, ADFS, Okta et OpenAthens) avec des options de configuration rapide pour les membres des fédérations *InCommon* et *UK Federation*, CAS, LDAP, OAuth 2, SIP2, SirsiDynix Symphony et Innovative Polaris.
b. Chaque compte doit être défini comme "régulier" ou "administrateur". Le rôle administrateur est réservé aux administrateurs de l'application en charge de la configuration. Un compte régulier peut se voir attribuer des droits élargis dans un planning donné, et devenir ainsi gestionnaire, pour :
    ○ créer, modifier et supprimer des quarts de travail ;
    ○ accéder aux statistiques du planning ;
    ○ gérer des demandes d'indisponibilité.
Les droits relatifs à l'accès aux statistiques et à la possibilité de gestion des demandes d'indisponibilité ne peuvent être attribués à un compte régulier que si celui-ci s'est déjà vu donner la possibilité de créer, modifier et supprimer des quarts de travail.
c. Afin de guider l'administrateur dans son travail de création et de configuration d'un planning, il existe un schéma de saisie, une suite d'opérations à réaliser via différents onglets.
d. Toutefois, pas de façon automatique comme c'est le cas pour les autres produits Springshare (*LibAnswers*, *LibCal* et *LibWizard*) pour lesquels il existe des procédures automatisables[15]. *LibStaffer* étant une solution réservée exclusivement au staff et aux agents en soutien, la seule façon de supprimer les données personnelles (prénom, nom et adresse email) consiste à ce qu'un administrateur supprime les comptes devenus inutiles ou les anonymise lui-même via la composante *LibApps* (tokenisation)[16].

---

[15] Voir par exemple *System Settings: Enable scrubbing of private patron data from LibWizard submissions* <https://ask.springshare.com/libwizard/faq/2501> (consulté le 2 mars 2022) et *Scrubbing Private Data in LibCal* <https://springyu.springshare.com/friendly.php?s=blocks/libcal/privacy-scrub> [accès après authentification] (consulté le 2 mars 2022)

11) L'interopérabilité avec d'autres logiciels

| Question | Réponse | Note |
|---|---|---|
| Le logiciel permet-il d'importer les heures de présence du personnel depuis un logiciel de gestion des ressources humaines (au format CSV ou dans un autre format) ? | Non | |
| Le logiciel permet-il d'importer les périodes de congés du personnel enregistrées dans un calendrier externe (au format iCalendar) ? | Oui | |
| Le logiciel permet-il d'importer des événements (absences, indisponibilités…) depuis l'agenda personnel des agents (au format iCalendar) ? | Oui | a |
| Le logiciel permet-il d'exporter l'emploi du temps personnel des agents vers un logiciel d'agenda (au format iCalendar) ? | Oui | |
| Peut-on accéder aux calendriers gérés par le logiciel via le protocole CalDAV (*Calendaring Extensions to WebDAV*) ? | Non | |

a. *LibStaffer* peut s'interfacer avec les calendriers *Outlook/Exchange* et *Google Calendar* et y ajouter de nouveaux quarts de travail assignés. De même, il pourra détecter un conflit en cas de nouvelle assignation qui tomberait en même temps qu'un événement déjà renseigné dans les calendriers *Outlook/Exchange* ou *Google Calendar*.

Signalons également que *LibStaffer* peut aussi s'interfacer avec "LibCal Appointments", une des composantes de la solution *LibCal*. "LibCal Appointments" permet de gérer la prise de rendez-vous avec du personnel de la bibliothèque afin de bénéficier d'un entretien ou d'un support particulier (tutorat…). Un rendez-vous pris via "LibCal Appointments" indiquera automatiquement l'agent comme indisponible dans son planning *LibStaffer*, sans pour autant le relever d'un quart de travail déjà planifié.

12) L'interface

| Question | Réponse | Note |
|---|---|---|
| *Le logiciel dispose-t-il d'une interface en français ? | Oui | a |
| *L'administrateur peut-il communiquer ou publier des news dans le logiciel ? | Oui | |
| *Le logiciel propose-t-il aux agents une vue de leurs assignations et services à venir (tableau de bord) ? | Oui | |

---

[16] En ce qui concerne le respect du Règlement général sur la protection des données (RGPD), Springshare annonce la parfaite conformité de ses produits (*GDPR Compliance* <https://springshare.com/gdpr.html> (consulté le 2 mars 2022)).

| | | |
|---|---|---|
| *En cas de modification ou de changement dans le planning, le logiciel permet-il de notifier les agents… | | |
| ● par email ? | Oui | |
| ● par SMS ? | Oui | b |
| ● via un autre canal de communication ? | Non | |
| *Le logiciel est-il couplé avec une app utilisable par les agents ? | Non | |
| *L'interface du logiciel répond-elle aux attentes du web réactif ? | Oui | |

  a. L'interface *LibStaffer* existe en anglais, en français et en espagnol. Cependant, à l'heure actuelle, la qualité de la traduction française peut parfois laisser à désirer[17].
  b. Toutefois, les possibilités de notification par SMS ne sont actuellement techniquement possibles qu'aux États-Unis et au Canada.

**Conclusions**

L'analyse conduite a pu déterminer qu'en termes de fonctionnalités disponibles, les réponses au cahier des charges étaient satisfaisantes. La petite vingtaine de questions supplémentaires adressées par les équipes de ULiège Library a permis en outre d'orienter l'outil vers des besoins réels de la bibliothèque. Les deux points faibles principaux dans le cas de l'ULiège sont certainement l'imperfection actuelle de certaines traductions françaises ainsi que l'impossibilité d'assigner simultanément et de façon récurrente un même agent à deux postes dans le même créneau horaire de la même journée. Dans ce dernier cas toutefois, une solution de contournement reste possible en cas de nécessité. Le fait que *LibStaffer* permette d'avoir une vue globale sur plusieurs plannings ou sur les tâches d'un agent affecté à différents plannings a en revanche été vivement apprécié par chaque utilisateur.

Aussi, après avoir testé les possibilités de l'outil pendant près de deux mois, avons-nous estimé que *LibStaffer* pouvait convenir à nos besoins en simplifiant la gestion du planning de l'accueil des implantations et des activités transversales. Nous avons donc décidé d'y souscrire pour une période de deux ans qui pourra être renouvelée si nous sommes satisfaits de la solution.

Logiciel proposé en mode SaaS, la configuration de base a été simple et rapide. Nous avons ensuite opté pour un déploiement graduel de *LibStaffer* sur une durée de 6 mois, implantation par implantation et service par service, pour permettre aux administrateurs de paramétrer progressivement les différents 21 plannings nécessaires, de présenter l'outil à chaque équipe individuellement et d'être en mesure d'encadrer chacune de celles-ci.

L'implémentation, entamée depuis mi-décembre 2021, offre déjà de la part des utilisateurs plusieurs pistes d'ajustement de l'outil (dont certaines font l'objet de demandes adressées à Springshare, comme par exemple l'assignation récurrente, dans un même créneau horaire, d'un même agent dans deux quarts de travail distincts) ou de solutions pour une meilleure utilisation propre à nos besoins. Il ne s'agit en effet pas de conformer nos désidératas à un système établi, mais bien de choisir un outil professionnel suffisamment développé pour qu'il soit apte à répondre au mieux à notre besoin : un planning simplifié, adapté à la fois à l'accueil et aux activités transversales, pour des membres du personnel, donc certains répartis sur plusieurs sites, avec des compétences et des horaires de travail différents.

---

[17] Exemples : *Shift Name* - Nom de décalage (au lieu de par exemple "Nom du quart de travail") ; *Open Shifts* - Ouvrir les quarts de travail (au lieu de par exemple "Quarts de travail disponibles") ; *Shift Timeline* - Décalage de la chronologie (au lieu de par exemple "Chronologie par quarts de travail"). Toutefois, le client francophone peut faire remonter au support technique de Springshare toute erreur de traduction et en proposer une meilleure au bénéfice de la communauté.

Plusieurs étapes d'un retour, à la fois global et détaillé, par les équipes utilisant *LibStaffer* sont prévues. L'expérimentation sur le terrain même, avec l'apport des nombreux collègues directement concernés, sera le meilleur moyen de dégager la conclusion d'un essai de deux années d'utilisation du logiciel de gestion de planning en bibliothèque *LibStaffer*.

## Bibliographie

- Becker-Redd, Kindra ; Lee, Kirsten ; Skelton, Caroline. Training Student Workers for Cross-Departmental Success in an Academic Library: A New Model, *Journal of Library Administration* [en ligne], 2018 (consulté le 2 mars 2022), vol. 58, n° 2, p. 153-165. <https://doi.org/10.1080/01930826.2017.1412711>.
- Dworak, Ellie. LibStaffer from Springshare. *The Charleston Advisor* [en ligne], January 2014 (consulté le 2 mars 2022), vol. 15, n° 3, p. 32-35. <https://doi.org/10.5260/chara.15.3.32>.
- Hughes, Sarah E. Scheduling Using a Web-Based Calendar: How Teamup Enhances Communication. *Public Services Quarterly* [en ligne], 2018 (consulté le 2 mars 2022), vol. 14, n° 4, p. 362-372 <https://doi.org/10.1080/15228959.2018.1518184>.
- Ishak, Yuyun Wirawati. Technology that enables, services that empower. *IFLA WLIC 2018: Reference and Information Services joint with Information Technology Satellite, Kuala Lumpur, Malaysia, August 23* [en ligne], 2018 (consulté le 2 mars 2022). <https://ink.library.smu.edu.sg/library_research/128>.
- Jankowski, Janiece. Successful Implementation of Six Sigma to Schedule Student Staffing for Circulation Service Desks, *Journal of Access Services* [en ligne], 2013 (consulté le 2 mars 2022), vol. 10, n° 4, p. 197-216. <https://doi.org/10.1080/15367967.2013.830930>.
- Lenepveu, Philippe ; Maisonneuve, Marc. *Les logiciels de gestion de planning de service pour les bibliothèques*. Éditions Klog, 2019. ISBN 979-10-92272-17-8.
- Springshare. *Get help with LibStaffer* [en ligne]. <https://ask.springshare.com/libstaffer/> (consulté le 2 mars 2022).
- Tang, Bixia H. Inviting student employees to work – a scheduling system built with free google applications, *Public Services Quarterly* [en ligne], 2020 (consulté le 2 mars 2022), vol. 16, n° 4, p. 254-264. <https://doi.org/10.1080/15228959.2020.1806178>.

**Auteurs**
François Renaville
Responsable des systèmes documentaires informatisés
Université de Liège. ULiège Library
Quartier Urbanistes, bât B63d
Traverse des Architectes, 5D
4000 Liège

francois.renaville@uliege.be
https://lib.uliege.be

Fabienne Prosmans
Responsable Service aux usagers et PIB, responsable scientifique Polytech
Université de Liège. ULiège Library
Quartier Polytech 1, bât. B52/4
Allée de la Découverte, 13A
4000 Liège
fprosmans@uliege.be
https://lib.uliege.be

Isabelle Gilles
Coordinatrice des ressources humaines et immobilières de ULiège Library
Université de Liège. ULiège Library
Quartier Urbanistes, bât B63d
Traverse des Architectes, 5D
4000 Liège
Isabelle.Gilles@uliege.be
https://lib.uliege.be

Mars 2022